\documentclass[12pt]{article}
\usepackage{color,soul}

\usepackage{amsmath,amssymb,amsthm}

\usepackage{hyperref}
\begin{document}
\begin{center}

{\bf FUNDAMENTAL SCALAR
FIELDS AND THE DARK SIDE\\ OF THE UNIVERSE}\vskip5pt
\vskip40pt{\footnotesize{EDUARD G. MYCHELKIN$^1$ and MAXIM A. MAKUKOV$^2$}
\vskip9pt
{\small
{\it Fesenkov Astrophysical Institute,\\ 050020, Almaty, Republic of Kazakhstan}
\vskip9pt
{$^1$mychelkin@aphi.kz, edmych@gmail.com\\ $^2$makukov@aphi.kz}}
}
\end{center}

\vskip40pt 
\begin{center}
	{\small{Essay written for the\\ Gravity Research Foundation 2015 Awards \\ (received an Honorable Mention) \\ Accepted in IJMPD }}
\end{center}

\vskip40pt 
	\begin{abstract}
		\noindent Starting with geometrical premises, we infer the existence of fundamental cosmological scalar fields. We then consider physically relevant situations in which spacetime metric is induced by one or, in general, by two scalar fields, in accord with the Papapetrou algorithm. The first of these fields, identified with dark energy, has exceedingly small but finite (subquantum) Hubble mass scale ($\approx$ $10^{-33}$ eV), and might be represented as a neutral superposition of quasi-static electric fields. The second field is identified with dark matter as an effectively scalar conglomerate composed of primordial neutrinos and antineutrinos in a special tachyonic state.
	\end{abstract}

\newpage

\hfil\break {\bf 1. Geometrical premises} \vskip10pt
\indent Unlike the $U(1)\times
SU(2)_{L}\times SU(3)$ Standard Model, gravity has no internal
symmetries but only those of space-time. In general, they might be
described by the spacetime deformation tensor $D_{\mu \nu}$ being
the Lie derivative $\mathcal{L_{\xi}}$ of the metric tensor
$g_{\mu\nu}$ with respect to some timelike field $\xi^{\mu}=\xi
u^{\mu}$ ($u^\mu u_\mu=1$, $\xi=\sqrt{\xi_{\mu}\xi^{\mu}}=\xi_{\mu} u^{\mu}$):
\begin{equation}
D_{\mu\nu}=\mathcal{L_{\xi}}g_{\mu\nu}\equiv\xi_{\mu;\nu}+\xi_{\nu;\mu}.
\label{DefTensor}
\end{equation}
The same $\xi^{\mu}$-field might be used to define the Riemann-Christoffel curvature tensor ${{R}^{\alpha}}_{\beta \gamma \delta }$:
\begin{equation}
{{\xi }^{\alpha }}{{R}_{\alpha \mu \nu \lambda }}={{\xi }_{\mu ;\nu \lambda }}-{{\xi }_{\mu ;\lambda \nu }}.
\label{CurvTensor}
\end{equation}
In the simplest case with $\xi=1$, a fundamental identity relating the Ricci tensor ${R}_{\alpha \beta}$ to the time-like $u$-congruence had been deduced from (\ref{CurvTensor}) by Ehlers \cite{Ehlers61} (in non-covariant form it was first obtained by Raychaudhuri \cite{Raych}):
\begin{equation}
{{R}_{\alpha \beta }}{{u}^{\alpha }}{{u}^{\beta }}=-\dot{\theta }+{{u}^{\beta }}{{u}^{\alpha }}_{;\beta \alpha }=-\dot{\theta }+{{\dot{u}}^{\alpha }}_{;\alpha }-{{u}_{\alpha ;\beta }}{{u}^{\beta ;\alpha }},
\label{EI}
\end{equation}
where $\theta ={{\nabla }_{\alpha }}{{u}^{\alpha }}\equiv {{u}^{\alpha }}_{;\alpha }$ is expansion, overdot denotes the time derivative, $\dot{\theta }={{u}^{\alpha }}{{\nabla }_{\alpha }}\theta$, and the tensor $u_{\alpha ; \beta}$ can be decomposed into irreducible components representing expansion, vorticity and shear \cite{LPPT}.\\ 
\indent This approach can be generalized for an arbitrary field $\xi^{\mu}$ in (\ref{DefTensor}) and (\ref{CurvTensor}). As a result, we deduce the generalized Ehlers identity relating the Ricci tensor not only to $u$-congruence, but also to the scalar field $\xi$ as the modulus of $\xi^\mu$:
\begin{equation}
R_{\alpha\beta}u^{\alpha}u^{\beta}=-\xi^{-1}\Box\xi
+u_{\alpha;\beta}u^{\alpha;\beta} +\xi^{-1}f_{\alpha}u^{\alpha},
\label{GenEI}
\end{equation}
$$ f^\mu \equiv
({D^\alpha}_\alpha)^{;\mu}-\tfrac{1}{2}{D^{\mu\alpha}}_{;\alpha}, \qquad
\Box\xi \equiv {\xi^{;\alpha}}_{;\alpha}.$$ 
\vskip5pt
\noindent For $\xi=1$ expression (\ref{GenEI}) reduces to (\ref{EI}), taking into account the relation between $u_{\alpha;\beta}u^{\alpha;\beta}$ and ${{u}_{\alpha ;\beta }}{{u}^{\beta ;\alpha }}$. The advantage of the form (\ref{GenEI}) is that it allows to consider such important characteristics of physical systems as redshift, temperature, potentials, etc. Following the paradigm of the correspondence between geometry and gravity, we conclude that the geometrical field $\xi$ implies the existence of a physical scalar counterpart $\phi$. In case of a Killing symmetry{\footnote{In this case $f^\mu=0$,
$u_{\alpha ; \beta}= -\dot{u}_\alpha u_\beta$  ($\dot{u}_\alpha$ is four-acceleration), and (\ref{GenEI}) reduces to the
Hawking-Ellis relation $R_{\mu\nu}u^{\mu}u^\nu=-\xi^{-1}h^{\alpha\beta}\xi_{;\alpha\beta}$ used for non-variational derivation of the Einstein equations \cite{HawEll}. In cases of conformal and space-conformal symmetries we have $D_{\mu\nu}=\Omega(x^{\mu}) g_{\mu\nu}$ \, and \, $D_{\mu\nu}=\Psi(x^{\mu}) h_{\mu\nu}$, correspondingly (with $h_{\mu\nu}=g_{\mu\nu}-u_{\mu}u_{\nu}$).}},
for which, by definition, $D_{\mu \nu}=0$, the modulus of the time-like Killing field is universally related to the metric as $\xi=\sqrt{g_{00}}$ \cite{LPPT}. At the same time, there exists a class of spherically symmetric metrics of the type $ds^2=Adt^2-Bd{\bf r}^2$ for which the condition $AB=1$ must hold, following from the principles of conformal symmetry \cite{Consoli} and relativistic quantum mechanics considerations \cite{Yilmaz82}. Thereby, we arrive at preferable metrics of the form
\begin{equation}
ds^2=\xi^{2}dt^2-\xi^{-2}d{\bf r}^2=e^{-2\phi}dt^2-e^{2\phi}d{\bf r}^2,
\label{PapaMetric1}
\end{equation}
where the Killing modulus is rescaled as $\xi=e^{-\phi}$, $c=1$, with positively defined $\phi$. Remarkably, this metric relation, derived here primarily from geometrical considerations, had been obtained by Papapetrou \cite{Papa} as the solution of Einstein's equations for the Killing case on the basis of ansatz  $g_{\mu\nu} =g_{\mu\nu} (\phi(x^\alpha))$, which implies that physical metrics can depend on coordinates only through some universal scalar $\phi$-field{\footnote {This implies that the energy-momentum tensor of this fundamental scalar field cannot be removed from the field equations, effectively ruling out any pure-vacuum solutions.}}.\\
\indent Papapetrou had also shown that in general case the metric might be induced by at most two independent background scalar fields, $g_{\mu\nu} =g_{\mu\nu} (\phi_1(x^\alpha), \phi_2(x^\alpha))$ \cite{Papa2}. This is justified in fact, given that the ten Einstein's equations might be reduced to at most two independent second-order equations if one excludes four zero-components representing first-order constraints and employs the four contracted Bianchi identities. Thus, at most two scalar fields might be used to close the initial system. We elaborate this idea to describe the global phenomena - dark energy (DE) and dark matter (DM) - as two independent fundamental scalar fields \cite{Mych1,mych09}.

\vskip10pt \hfil\break {\bf 2. Dark energy}
\vskip10pt We begin with the Papapetrou solution (\ref{PapaMetric1})  of the Einstein equations for a minimal spherically-symmetric antiscalar field $\phi$:
\begin{equation}
ds^{2} = e^{ - 2\phi} dt^{2} - e^{2\phi} (dr^{2} + r^{2}d\Omega
^{2}) = e^{ - 2GM / r}dt^{2} - e^{2GM / r}(dr^{2} + r^{2}d\Omega
^{2}).
\label{PapaMetric2}
\end{equation}
Antiscalarity implies that the sign of the energy-momentum tensor of the scalar field in the field equations is opposite to that of ordinary matter: $G_{\mu\nu}=\kappa T_{\mu\nu}=\kappa
(-T^{scalar}_{\mu\nu}+T^{matter}_{\mu\nu}+...)$. As shown in \cite{MychMak}, this follows from the requirement of thermodynamic stability of scalar fields\footnote{The issue of the sign of $T^{scalar}_{\mu\nu}$ was pointed out by Yilmaz \cite{Yilmaz58}, and also discussed in \cite{Ell}.}. Solution (\ref{PapaMetric2}) contains the observed ``crucial effects" and leads to the correct formulae for lensing. It does not contain black holes (see footnote 2), but for compact objects
with the scale of the order of the gravitational radius $2GM/c^{2}$ we get, up to constant factors, familiar results of black hole thermodynamics. The effective background field $\phi$ represents neutral superposition $\phi=(\phi^{+}+\phi^{-})/\sqrt{2}$ of primordial quasi-static electric
fields manifesting itself only through gravity.{\footnote{This is justified by the fact that both the Einstein-Maxwell and the minimal antiscalar field equations in quasi-static limit reduce to the same antiscalar form \cite{MychMak,Zhuravlev,Marsh} (up to the `balance condition' between charges and masses, $q^2=G m^2$, or between the corresponding densities).} Heuristically, all particles and fields (except gravity) might be represented as excited states of the primordial $\phi$-field, since any particle and antiparticle can annihilate into photons, and a photon might then be transformed kinematically (via Doppler effect) into quasi-static state equivalent to the $\phi$-field. And the entire reverse process, from the quasi-static $\phi$-field to particles, is, in principle, also conceivable.\\
\indent Considering cosmological scales in space-flat Friedmannian metric (corresponding to space-conformal symmetry, see footnote 1), we supplement the given scalar field with mass- and cosmological terms but with no any additional self-interaction potentials:
\begin{equation}
T^{scalar}_{\mu\nu}=\frac{1}{4\pi}[\phi_{,\mu} \phi_{,\nu}
-\frac{1}{2}g_{\mu\nu}
      (\phi_{,\alpha} \phi^{,\alpha} -\mu^2 \phi^2 -
      \Lambda/G)],
\label{ScalarTensor}
\end{equation}
$$
G_{\mu\nu}=-\kappa T^{scalar}_{\mu\nu},
$$
 where $\mu^{2}=-m^{2}<0$ for quasi-static fields imitating instant action at a distance. The negative sign of $T^{scalar}_{\mu\nu}$ is due to antiscalarity, and for now we neglect here $T^{matter}_{\mu\nu}$ and other components. Then, under the general integrability condition \cite{Maciejewski} $\Lambda = -(2/3)\mu^2$ we have $|\mu|=m \approx
10^{-33}$ eV, and solving also the corresponding Klein-Gordon equation, we ultimately arrive at the Gaussian cosmological solution for the primordial universe \cite{mych09} (see also \cite{Szekeres,Siqueira, Holten}), which allows to identify this antiscalar field with dark energy background:
\begin{equation}
ds^2=dt^2-a^2(\phi)(dr^2+r^2 d\Omega^2),\,\,\,\,
a^2(\phi)=e^{-\phi^2}=e^{-\Lambda(t-t_{0})^2}.
\label{SzekeresMetric}
\end{equation}
Then, the equation of state for DE with effective variable parameter $w$ follows \cite{mych09}:
$p=w\varepsilon, \,\, \varepsilon=T_{0}^{0}, \,\, p=-T_{i}^{i}/3$,
\begin{equation}
w=\frac{1-3\Lambda(t-t_{0})^2 /2}{3\Lambda(t-t_{0})^2 /2}.
\label{EOS}
\end{equation}
Asymptotically at $t\rightarrow\infty$ it goes to the de Sitter
state with $w=-1$. When $w<-1/3$ we get from (\ref{EOS}) the accelerated phase for the universe evolution. After certain time $t_{0}$ expansion always
changes to contraction (even in the absence of other matter). As follows from (\ref{EOS}), in this approach phantom fields with
$w<-1$ can never arise.\\
\indent Now, for a central mass $M$ on DE background we get the solution
\begin{equation}
ds^{2} = e^{ - 2GM / r}dt^{2} - e^{2GM /
r}e^{-\Lambda(t-t_{0})^2}(dr^{2} + r^{2}d\Omega ^{2}),
\label{uni}
\end{equation}
which represents the unification of local (\ref{PapaMetric2}) and cosmological (\ref{SzekeresMetric}) effects into a single metric of the type:
\begin{equation}
ds^{2} = e^{ - 2\phi} dt^{2} - e^{2\phi} a^2(t)(dr^{2} +
r^{2}d\Omega ^{2}).
\label{UnifiedMetric}
\end{equation}
In linear approximation, (\ref{UnifiedMetric}) coincides with the widely used perturbative solution in Newtonian gauge describing the linear stage of structure formation in the universe:
\begin{equation}
ds^{2} =(1-2\phi) dt^{2} - (1+2\phi) a^2(t)(dr^{2} +
r^{2}d\Omega ^{2}).
\label{PertSol}
\end{equation}
If DM can be described by another scalar field $\phi_{DM}$ (see below), then it might be included into $\phi$ in (\ref{UnifiedMetric}) and (\ref{PertSol}):
\begin{equation}
\phi(r)=\phi_{Newton}(r)+\phi_{DM}(r).
\label{FullPot}
\end{equation}
\indent Note that since the values of mass- and $\Lambda$-terms of DE are of the same order of magnitude, they should always enter equations only together.

\vskip5pt \hfil\break {\bfseries 3. Dark matter} \vskip10pt Given a number of inconsistencies faced by traditional cold, warm and hot dark matter models, we develop an alternative approach to DM as a gravitating tachyonic background. This is, in a sense, the revival of the historically first DM model based on neutrinos \cite{Cowsik}, but with a further pivotal assumption that neutrinos might exist in tachyonic almost sterile state. Such assumption is consistent both with the principles of special relativity (see, e.g., \cite{Geroch,HillCox}) and with a number of observational data \cite{Ehrlich}. The possibility that DM might have a tachyonic component was first considered in \cite{Davies}, with the conclusion that tachyons could affect the universe evolution until now only if their mass is sufficiently small, $m \lesssim 0.1$ eV. This is exactly the current upper bound for neutrino masses. \\
\indent However, tachyonity is typically related to such problems as violation of causality and/or unitarity, instability of tachyonic modes, negative energies, violation of the Pauli principle, the lack of a generally accepted scheme of (second) quantization for tachyons, difficulties with the classification of tachyons by irreducible representations of the Poincar\'e group, etc. As a first  step to avoid these problems we propose an alternative group-theoretical algorithm for the adequate description of tachyons. \\
\indent The essential point is that the so-called `tachyonic representations of the Lorentz group' should be replaced by approach based on a distinct `tachyon Lorentz group' which does not employ imaginary masses and negative energies, and, as a consequence, does not lead to violation of causality and unitarity. To this end, we upgrade the algorithm first proposed in \cite{Recami}, to make it the inversion of particles' {\it squared} velocities (in units of $c$): 
\begin{equation}
v^2=1/u^2 \quad \rightleftarrows \quad u^2=1/v^2, \qquad 0\le v<1, \qquad 1<u<\infty.
\label{inversion}
\end{equation}
Such mapping leads to the distinct tachyon Lorentz group with different (superluminal) parametrization which acts, strictly speaking, on another Minkowski space with the opposite-signature metric. Then, in the four-momentum space one gets the correspondence between bradyon and tachyon dispersion relations:
\begin{equation}
E^{2} (v)-p^{2} (v)=m^{2} \quad \rightleftarrows \quad \tilde{p}^{2} (u)-\tilde{E}^{2}(u)=m^{2},
\end{equation}
where bradyon and tachyon energies are, correspondingly:
$$E(v)=m/\sqrt{(1-v^2)}, \qquad \quad \tilde{E}(u)=m/\sqrt{({{u}^{2}}-1)}.$$
Here masses and energies are put to be always real and positively defined in both four-momentum spaces (each with its own symmetry group). Furthermore, by changing the sign in both inversion formulas in (\ref{inversion}), one arrives at two adjacent subgroups of the full Euclidean (instanton) Lorentz group acting on spacetime and four-momentum space with the signature (++++) \cite{Ramond}. All mentioned Lorentz-type groups describing bradyons, tachyons and instantons might be combined into the ``Lorentz groupoid" acting on the corresponding tangent spacetime and the four-momentum space.\\
\indent Accordingly, for each of these groups the Dirac $\gamma$-matrices should satisfy their own anticommutation relations of the Clifford algebra, $\{\gamma^{\mu},\gamma^{\nu}\}=2g^{\mu\nu}$, defined by corresponding metric tensors $g^{\mu\nu}$. Applying the Lorentz groupoid to spinors, one finds that to maintain covariance with respect to the corresponding group, only for customary (bradyon) Lorentz group the Hermitian conjugation should be replaced with Dirac conjugation (via multiplication by Hermitian matrix $\gamma_0$,  $\bar{\psi }=\psi ^{\dag } \gamma _{0} $). However, for the Euclidean Lorentz group only Hermitian conjugation is necessary for covariance \cite{Ramond}, and the same is true for the tachyon Lorentz group, as well as for the Galilean group (which are also compatible with the existence of superluminal particles).\\
\indent Operating with Hermitian conjugation of wave functions, we obtain the mass-term well-defined separately for neutrinos and antineutrinos (this is impossible with Dirac conjugation \cite{Zee}). As a result, superposition of the squares of free tachyon neutrino $\nu $ and antineutrino $\bar{\nu }$ spinor fields might be represented as a scalar conglomerate\footnote{We use the term conglomerate, as opposed to condensate (not quite appropriately used in \cite{Mych1}), because tachyons violate the Pauli principle and have no definite spin values (specified according to the Casimir invariants of the Poincar\'e group \cite{Hughes,Bekaert}), so we retain them unquantized.}: 
\begin{equation}
\Phi =\psi ^{\dag } \psi =\nu ^{2} +\bar{\nu }^{2}, 
\label{congl}
\end{equation}
 with $\psi =\nu +i\bar{\nu }$. Such effectively scalar and unquantized tachyon field $\Phi =\phi_{DM}(r)$ in the Einstein and Klein-Gordon equations (see (\ref{FullPot})) leads to the interpretation of dark matter phenomenon as the primordial gravitating neutrino-antineutrino conglomerate \cite{mych09}.\\
\indent While differing in physical interpretations, in operational sense the approaches employing imaginary and real masses are equivalent (at least if one sets quantum aspects of tachyonic physics aside). E.g., if one writes the following tachyonic version of the Dirac equation within the traditional approach:
\begin{equation}
\left(i\gamma ^{\mu } \partial _{\mu } -\Gamma m\right)\psi =0, 
\label{TachDirac}
\end{equation}
where $\Gamma =\gamma _{0} \gamma _{5}$ serves as the matrix equivalent of imaginary unit ($\Gamma ^{2}=-\bf{1}$ in any representation of $\gamma$-matrices), then, unlike customary Dirac equation, it might be split into two independent equations (in terms of the momentum operators and the Pauli matrices): 
\begin{equation}
\left(p_{0} -\vec{\sigma }\vec{p}-m\right)\psi _{L} =0, \qquad
\left(p_{0} +\vec{\sigma }\vec{p}+m\right)\psi _{R} =0, 
\label{SplitDirac}
\end{equation}
separately for left ($\psi _{L}=\nu$) neutrinos and right ($\psi _{R}=\bar{\nu}$) antineutrinos\footnote{This implies that tachyonic neutrinos cannot be of Majorana type, in agreement with (as yet) null experimental results \cite{EXO}.}. Such splitting also makes it possible to build a scalar conglomerate out of neutrinos and antineutrinos.\\
\indent On the scales of galaxies and clusters, tachyonic scalar neutrino-antineutrino conglomerate is not subject to physical constraints on the value of its density (as opposed to condensate \cite{TremaineGunn}). This density might be considered as a free parameter and put to be equal to the observed DM density. In quasi-stationary approximation this practically sterile (especially for low energies) neutrino background is distributed all over the universe and produces somewhat denser regions (``smoothed halos" \cite{MakukRG15}) around galaxies and clusters. Tachyonic neutrinos can possess almost stiff equation of state, $p\approx\varepsilon$, which might be related to the isothermal sphere profile, $\varepsilon\sim r^{-2}$ (see in \cite{Luka}). The latter produces the logarithmic-type potential and leads to the observed flat rotation curves for galaxies.\\
\indent The structure formation in the universe is outlined tentatively as follows. Galaxies begin to develop in accord with the standard Jeans instability of baryonic matter. Due to extremely weak interactions with tachyonic neutrinos, small-scale fluctuations of baryonic matter are not washed out. Their subsequent growth then accelerates due to the permanent background of the surrounding neutrino-antineutrino conglomerate. In turn, only large-scale fluctuations survive in the neutrino conglomerate itself. Therefore, as opposed to bottom-up and top-down scenarios, in this case structures at smallest and largest scales develop more or less simultaneously, and observations of very large structures like the Huge-LQG and Hercules-Corona Borealis Great Wall should come as less surprise. \\
\indent It should be noted that the existence of the primordial tachyon neutrino DM background considered here does not imply by itself that the secondary (ordinary) neutrinos produced at cosmological temperatures of about a few MeV from lepton annihilations must be of tachyonic nature as well. It is possibile that the rest of neutrinos are produced in bradyonic (subluminal) states, and if so, the standard cosmological scenario (maintaining the local thermodynamic equilibrium) is not altered essentially but supplemented with the effects of the primordial background of tachyonic neutrinos. The kinematic distinction between high-energy tachyonic and bradyonic neutrinos is extremely small, so the ultimate decision is up to future experiments.

\vskip5pt \hfil\break {\bfseries 4. Conclusion} \vskip10pt
\indent Unlike conventional approaches we do not take any
\emph{ad hoc} scalar fields to describe DE and DM but \emph{deduce} the first
fundamental scalar field from general principles, and only at the last step identify it with DE. This field is represented by neutral superposition of ubiquitous quasi-static electric fields which, in accord with Schwinger's conjecture, should possess their own carriers \cite{Schwinger}. These, in our case, have extremely small but finite mass of about $10^{-33}$ eV, and might be dubbed `statons'. \\
\indent For the second scalar DM field we propose neutrinos as its constituents, provided that they are tachyons. The discovered effect of neutrino oscillations implies that neutrinos have non-zero masses and thus cannot travel at the speed of light. On the other hand, according to experiments on parity violation in weak interactions, all neutrinos are left (and antineutrinos are right). From heuristic consideration, if neutrino velocities were less than the speed of light, in some reference frames neutrino helicity would swap to the opposite. As this has never been observed (within the accuracy of experiments), the natural conclusion is that neutrino velocities should be greater than the speed of light. Thus, tachyonity of neutrinos can, in principle, be considered as a consequence of the chiral invariance, rather than an \emph{ad hoc} assumption. If neutrinos are indeed tachyons, there is no necessity to invoke hypothetical right-handed sterile neutrinos to explain the origin of neutrino mass states.\\
\indent On the whole, in this approach we envisage two well-defined background fields, DE and DM, with comparable densities but drastically different mass scales, making these fields practically non-interacting during almost all of the universe evolution. In this model, there is no necessity to introduce any \emph{ad hoc} solutions such as self-interaction potentials (except for the mass-term), exotic particles unfamiliar to present-day experiments, and modifications of general relativity, to account for the dark side of the universe.

%\pagebreak


\begin{thebibliography}{}
{\footnotesize

\bibitem{Ehlers61}
{Ehlers J. Beitr{\"a}ge zur relativistischen Mechanik kontinuierlicher Medien. {\it Akad. Wiss. Lit. Mainz, Abhandl. Math.-Nat. Kl.} {\bf 11} (1961) 793-836.\\ \footnotesize{ English translation is available in {\it Gen. Rel. Grav.} {\bf 25} (1993) 1225-1266.}}

\bibitem{Raych}
{Raychaudhuri A. Relativistic cosmology. I. {\it Phys. Rev.} {\bf 98} (1955) 1123-1126.}

\bibitem{LPPT}
{Lightman A.P., Press W.H., Price R.H., Teukolsky S.A. {\it Problem Book in Relativity and Gravitation}. Princeton University Press (1975).}

\bibitem{HawEll}
{Hawking S.W., Ellis G.F.R. {\it The Large Scale Structure of Space-Time}. Cambridge University Press (1973).}

\bibitem{Consoli}
{Consoli M. Newtonian gravity from the Higgs field: the sublimation of aether. \href{http://arxiv.org/abs/hep-ph/0109215}{arXiv:hep-ph/0109215}.}

\bibitem{Yilmaz82}
{Yilmaz H. Relativity and quantum mechanics. {\it Int. J. Theor. Phys.} {\bf 21} (1982) 871-902.}

\bibitem{Papa}
{Papapetrou A. Eine Theorie des Gravitationsfeldes mit einer Feldfunktion. {\it Z. Phys.} {\bf 139} (1954) 518-532.}

\bibitem{Papa2}
{Papapetrou A. Eine neue Theorie des Gravitationsfeldes. I. {\it Math. Nachr.} {\bf 12} (1954) 129-141.}

\bibitem{Mych1}
{Mychelkin E.G. Inevitability of antiscalar gravity. In {\it Proceedings of International Scientific Meeting 'Number, Time, Relativity'}, ed. by Pavlov D.G., Asanov G.S. (Moscow, Russia, 2004), p. 47.
	(\href{http://hypercomplex.xpsweb.com/articles/194/en/pdf/sbornik.pdf}{http://hypercomplex.xpsweb.com/articles/194/en/pdf/sbornik.pdf})}.

\bibitem{mych09}
{Mychelkin E.G. Two fundamental fields: Identification of dark energy and dark matter. In {\it Proceedings of the MG12 Meeting on General Relativity, Part C}, ed. by Damour T., Jantzen R.T., Ruffini, R. (Paris, France, 2009), p. 1861. doi: \href{http://dx.doi.org/10.1142/9789814374552_0346}{10.1142/9789814374552\_0346}}

\bibitem{MychMak}
{Mychelkin E.G., Makukov M.A. Antiscalar cosmological background. \href{http://arxiv.org/abs/1506.03608}{arXiv:1506.03608 [gr-qc]}.}

\bibitem{Yilmaz58}
{Yilmaz H. New approach to general relativity. {\it Phys. Rev.} {\bf 111} (1958) 1417-1426.}

\bibitem{Ell}
{Ellis H.G. Ether flow through a drainhole: A particle model in general relativity. {\it J. Math. Phys.} {\bf 14} (1973) 104-118.}

\bibitem{Zhuravlev}
{Zhuravlev V.M., Kornilov D.A. Effect of latent mass in inhomogeneous cosmological model with perfect fluid and self-acting scalar field. {\it Grav. Cosmol.} {\bf 5} (1999) 325-328. \href{http://arxiv.org/abs/gr-qc/0002082}{arXiv:gr-qc/0002082}.}

\bibitem{Marsh}
{Marsh G.E. Charge, geometry, and effective mass. {\it Found. Phys.} {\bf 38} (2008) 293-300.\href{http://arxiv.org/abs/0708.1958}{arXiv:0708.1958 [gr-qc]}}

\bibitem{Maciejewski}
{Maciejewski A.J., Przybylska M., Stachowiak T., Szyd\l{}owski M. Global dynamics of cosmological scalar fields - Part II. \href{http://arxiv.org/abs/gr-qc/0703031}{arXiv: gr-qc/0703031}.}

\bibitem{Szekeres}
{Szekeres P. Cosmological singularities. In {\it Singularities, Black Holes and Cosmic Censorship}, ed. by Joshi P.S. (Pune, India, 1996), p. 55.}

\bibitem{Siqueira}
{de Siqueira A.C.V.V. A possible origin of dark matter, dark energy, and particle-antiparticle asymmetry. \href{http://arxiv.org/abs/1009.6193}{arXiv:1009.6193 [gr-qc]}.}

\bibitem{Holten}
{van Holten J.W. On single scalar field cosmology. \href{http://arxiv.org/abs/1301.1174}{arXiv:1301.1174 [gr-qc]}.}

\bibitem{Cowsik}
{Cowsik R., McClelland J. Gravity of neutrinos of nonzero mass in astrophysics. {\it Astrophys. J.} {\bf 180} (1973) 7-10.}

\bibitem{Geroch}
{Geroch R. Faster than light? {\it Advances in Lorentzian Geometry, AMS/IP Stud. Adv. Math.} {\bf 49} (2011) 59-69. \href{http://arxiv.org/abs/1005.1614}{arXiv:1005.1614 [gr-qc]}.}

\bibitem{HillCox}
{Hill J.M., Cox B.J. Einstein's special relativity beyond the speed of light. {\it Proc. R. Soc. A} {\bf 468} (2012) 4174-4192.}

\bibitem{Ehrlich}
{Ehrlich R. Six observations consistent with the electron neutrino being a $m^2=-0.11 \pm 0.02$ eV$^2$ tachyon. {\it Astropart. Phys.} {\bf 66} (2015) 11-17.}

\bibitem{Davies}
{Davies P.C.W. Tachyonic dark matter. {\it Int. J. Theor. Phys.} {\bf 43} (2004) 141-149. \href{http://arxiv.org/abs/astro-ph/0403048}{arXiv:astro-ph/0403048}.}

\bibitem{Recami}
{Pavsic M., Recami E. A new unified approach to bradyons and tachyons by conformal transformations. {\it Lett. Nuovo Cimento} {\bf 19} (1977) 273-278.}

\bibitem{Ramond}
{Ramond P. {\it Field Theory: A Modern Primer}. Benjamin Cummings, San Francisco (1981).}

\bibitem{Zee}
{Zee A. {\it Quantum Field Theory in a Nutshell}. Princeton University Press (2010).}

\bibitem{Hughes}
{Hughes R.J., Stephenson Jr. G.J. Against tachyonic neutrinos. {\it Phys. Lett. B} {\bf 244} (1990) 95-100.}

\bibitem{Bekaert}
{Bekaert X., Boulanger N. The unitary representations of the Poincar\'e group in any spacetime dimension. \href{http://arxiv.org/abs/hep-th/0611263}{arXiv:hep-th/0611263}.}

\bibitem{EXO}
{The EXO-200 Collaboration. Search for Majorana neutrinos with the first two years of EXO-200 data. {\it Nature} {\bf 510} (2014) 229-234.}

\bibitem{TremaineGunn}
{Tremaine S., Gunn J.E. Dynamical role of light neutral leptons in cosmology. {\it Phys. Rev. Lett.} {\bf 42} (1979) 407-410.}

\bibitem{MakukRG15}
{Makukov M.A., Mychelkin E.G., Saveliev V.L. Smoothed isothermal profile for tachyon neutrino dark matter. In {\it RUSGRAV-15 International Conference on Gravitation, Cosmology and Astrophysics} (Kazan, Russia, 2014), p. 184.}

\bibitem{Luka}
{Luk\'{a}cs A. Neutron stars with orbiting light? {\it Astron. Nachr.} {\bf 310} (1989) 49-59.} 

\bibitem{Schwinger}
{Schwinger J.S. {\it Particles, Sources and Fields}. Addison-Wesley, Boston (1973).}
}

\end{thebibliography}
\end{document}